\documentstyle[twocolumn,aps,prl,floats,epsf]{revtex}

\begin{document}

\draft

\wideabs{

\title{Measurement of the Light Antiquark Flavor Asymmetry in the
       Nucleon Sea}

\author{
E.A.~Hawker$^j$, 
T.C.~Awes$^i$,
M.E.~Beddo$^h$, 
C.N.~Brown$^c$, 
J.D.~Bush$^a$,
T.A.~Carey$^f$, 
T.H.~Chang$^h$,
W.E.~Cooper$^c$,
C.A.~Gagliardi$^j$,
G.T.~Garvey$^f$, 
D.F.~Geesaman$^b$, 
X.C.~He$^d$,
L.D.~Isenhower$^a$,
S.B.~Kaufman$^b$, 
D.M.~Kaplan$^e$, 
P.N.~Kirk$^g$, 
D.D.~Koetke$^k$, 
G.~Kyle$^h$,
D.M.~Lee$^f$,
W.M.~Lee$^d$, 
M.J.~Leitch$^f$, 
N.~Makins$^b$\cite{byline1}, 
P.L.~McGaughey$^f$, 
J.M.~Moss$^f$,
B.A.~Mueller$^b$,
P.M.~Nord$^k$,
B.K.~Park$^f$, 
V.~Papavassiliou$^h$, 
J.C.~Peng$^f$, 
G.~Petitt$^d$, 
P.E.~Reimer$^f$,
M.E.~Sadler$^a$,
J.~Selden$^h$, 
P.W.~Stankus$^i$, 
W.E.~Sondheim$^f$, 
T.N.~Thompson$^f$, 
R.S.~Towell$^a$\cite{byline2},
R.E.~Tribble$^j$,
M.A.~Vasiliev$^j$\cite{byline3}, 
Y.C.~Wang$^g$, 
Z.F.~Wang$^g$, 
J.C.~Webb$^h$, 
J.L.~Willis$^a$,
D.K.~Wise$^a$,
G.R.~Young$^i$\\ \vspace*{9pt}
(FNAL E866/NuSea Collaboration)\\ \vspace*{9pt}
}
\address{
$^a$Abilene Christian University, Abilene, TX 79699\\
$^b$Argonne National Laboratory, Argonne, IL 60439\\
$^c$Fermi National Accelerator Laboratory, Batavia, IL 60510\\
$^d$Georgia State University, Atlanta, GA 30303\\
$^e$Illinois Institute of Technology, Chicago, IL  60616\\
$^f$Los Alamos National Laboratory, Los Alamos, NM 87545\\
$^g$Louisiana State University, Baton Rouge, LA 70803\\
$^h$New Mexico State University, Las Cruces, NM, 88003\\
$^i$Oak Ridge National Laboratory, Oak Ridge, TN 37831\\
$^j$Texas A \& M University, College Station, TX 77843\\
$^k$Valparaiso University, Valparaiso, IN 46383
}
\date{\today}

\maketitle

\begin{abstract}
A precise measurement of the ratio of Drell-Yan yields from
an 800~GeV/c proton beam incident on hydrogen and deuterium targets is
reported.  Over 140,000 Drell-Yan muon pairs with dimuon mass
$M_{\mu^+\mu^-}\ge 4.5$~GeV/c$^2$ were recorded.  From these data, the
ratio of anti-down ($\bar{d}$) to anti-up ($\bar{u}$) quark
distributions in the proton sea is determined over a wide range in
Bjorken-$x$.  A strong $x$ dependence is observed in the ratio
$\bar{d}/\bar{u}$, showing substantial enhancement of $\bar{d}$ with
respect to $\bar{u}$ for $x<0.2$. This result is in fair agreement
with recent parton distribution parameterizations of the sea.  For
$x>0.2$, the observed $\bar{d}/\bar{u}$ ratio is much nearer unity
than given by the parameterizations.
\end{abstract}

} 


No known symmetry requires equality of the $\bar{d}$ and $\bar{u}$
distributions in the proton.  Until recently it had been generally
assumed that $\bar{d}(x)=\bar{u}(x)$ for lack of experimental evidence
to the contrary.  This assumption may be evaluated by use of the
expression
\begin{equation}
\int_{0}^{1}\left[F_{2}^{p}(x)-F_{2}^{n}(x)\right]
            \frac{\textstyle dx}{x} = 
\frac{1}{3} - \frac{2}{3} \int_{0}^{1}\left[\bar{d}_{p}(x)-
               \bar{u}_{p}(x)\right]{\text dx}.
\label{eq:1}
\end{equation}
Here $F_{2}^{p}(x)$ and $F_{2}^{n}(x)$ are the proton and neutron
inelastic structure functions, and $\bar{d}_{p}(x)$ and
$\bar{u}_{p}(x)$ are the anti-down and anti-up quark distributions in
the proton sea as a function of Bjorken-$x$\@.  Equation~\ref{eq:1}
requires the assumption of charge symmetry between the proton and
neutron (i.e. $u_p = d_n$, $\bar{u}_p = \bar{d}_n$, etc.).  If the
nucleon sea is flavor symmetric in the light quarks, the value of the
integral on the left is $1/3$, a result referred to as the Gottfried
Sum Rule (GSR)~\cite{gottfried}. In 1991 the New Muon Collaboration
(NMC) at CERN presented evidence that the GSR is violated, based on
deep inelastic muon scattering data from hydrogen ($p$) and deuterium
($d$). They reported a final value of
$\int_{0}^{1}\left[F_{2}^{p}(x)-F_{2}^{n}(x)\right]\frac{dx}{x} =
0.235 \pm 0.026$~\cite{amaudruz}, which implies that
\begin{equation}
\int_{0}^{1}\left[\bar{d}_{p}(x)-\bar{u}_{p}(x)\right]dx
   = 0.147\pm 0.039,
\label{eq:2}
\end{equation}
a considerable excess of $\bar{d}_{p}$ relative to $\bar{u}_{p}$.
This result has been adopted in the most current parameterizations of
the parton distributions in the nucleon~\cite{lai,martin}.

Following publication of the NMC result, the use of the Drell-Yan
process \cite{drell} was suggested \cite{ellis} as a means by which
the light antiquark content of the proton could be more directly
probed.  This was first done by the Fermi\-lab E772 collaboration.
They compared the production of Drell-Yan muon pairs from isoscalar
targets to that from a neutron rich target.  This measurement sets
constraints on the non-equality of $\bar{u}$ and $\bar{d}$ in the
range $0.04\le x\le 0.27$~\cite{mcgaughey}.  Later, the CERN
experiment NA51~\cite{baldit} carried out a comparison of the
Drell-Yan muon pair yield from hydrogen and deuterium at a single
value of $x$ using a 450~GeV/c proton beam and found
\begin{equation}
\left. \frac{\bar{u}_{p}}{\bar{d}_{p}} \right|_{\langle x \rangle=0.18}
 = 0.51\pm 0.04\pm 0.05.
\label{eq:3}
\end{equation}
A recent review by Kumano \cite{kumano} presents an extensive
discussion of the existing literature on the flavor asymmetry of the
antiquark sea.

Fermilab Experiment 866 (E866) measured the Drell-Yan muon pair yield
from 800~GeV/c proton bombardment of liquid deuterium and hydrogen
targets. From these data, $\bar{d}/\bar{u}$ and $\bar{d}-\bar{u}$ in
the proton over the range $0.020 < x < 0.345$ are extracted. A
significant difference between the $\bar{d}$ and $\bar{u}$
distributions is found.

E866 used a 3-dipole magnet spectrometer~\cite{moreno} employed in
previous experiments (E605, E772, and E789), modified by the addition
of new drift chambers and hodoscopes with larger acceptance at the
first tracking station.  Other improvements to the spectrometer
included a programmable trigger system~\cite{gagliardi} and a
VME-based data acquisition system.  An 800~GeV/c extracted proton beam
with up to $2\times 10^{12}$ protons per 20~s spill bombarded one of
three identical 50.8~cm long cylindrical target flasks containing
either liquid hydrogen, liquid deuterium or vacuum.  After passing
through the target, the remaining beam was intercepted by a copper
beam dump located inside the second dipole magnet.  The beam dump was
followed by a 13.4 interaction length absorber wall of copper, carbon
and polyethylene which blocked the entire aperture of the magnet.
This absorber wall removed hadrons produced in the target and the
dump, ensuring that only muons traversed the spectrometer's detectors.
The detection system consisted of four tracking stations and a
momentum analyzing magnet.  The spectrometer's acceptance as a
function of $p_T$, the transverse momentum of the dimuon pair, was
reasonable to 3.0~GeV/c, with some acceptance to 5.0~GeV/c.

The targets alternated between hydrogen and deuterium every five beam
spills with a single spill collected on the empty flask at each target
change.  The targets were 3.54 and 8.14~g/cm$^2$ thick, corresponding
to 7 and 15\% of an interaction length for hydrogen and deuterium,
respectively.  Beam intensity was monitored by secondary-emission
detectors, an ion chamber and quarter-wave RF cavities.  Two
four-element scintillator telescopes viewing the targets at nearly
$90^{\circ}$ monitored the luminosity, beam duty factor and data
acquisition live time.  The trigger required a pair of triple
hodoscope coincidences having the topology of a muon pair from the
target.  Typically 70 triggers per second were recorded with an
electronic live time above 98\%\@. An integrated flux of $1.3\times
10^{17}$ protons was delivered.

Over 330,000 Drell-Yan events were recorded, using three different
spectrometer settings which focused low, intermediate and high mass
muon pairs.  The data collected with the low and intermediate mass
settings have systematic effects of a few percent which require
additional study.  The data from the high mass setting are relatively
free from these effects due to the greatly reduced rates in the
tracking chambers.  Therefore, this Letter presents only the results
from the high mass setting, with over 140,000 Drell-Yan events.

In calculating the Drell-Yan yields, a small correction (averaging
0.2\%) for random coincidences between two unrelated, oppositely
charged muons was made.  This correction was evaluated by studying the
observed rates of same charge muon pairs. A background rate of
approximately 8\% from the target flask and beam line windows was
measured using the evacuated flask and was subtracted.  Given the
identical geometry of the deuterium and hydrogen targets and the
slight difference in the average interaction point due to beam
attenuation in the target, acceptance differences between the two
targets were shown by a Monte Carlo simulation of the spectrometer to
be very small.  The measured cross section ratio was corrected for
differences in beam attenuation in the targets, target density and a
small hydrogen contamination in the deuterium target.  Additionally,
the high luminosity caused a small, rate dependent inefficiency.  This
led to a correction of 1.2\% in the cross section ratio.  The
systematic error in the ratio of yields from the two targets is
dominated by the uncertainties in the rate dependence ($\pm$0.6\%),
hydrogen contamination in the deuterium target ($\pm$0.2\%) and beam
attenuation ($\pm$0.2\%)\@.  All other contributions, including the
production of muon pairs from secondary hadron reinteraction, are
small.  The total systematic error in the cross section ratio is less
than $\pm$1\%.

\begin{figure}
  \begin{center}
    \mbox{\epsffile{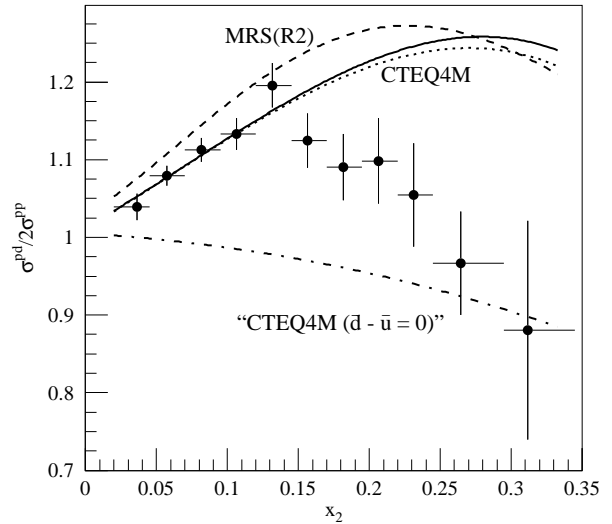}}
  \end{center}

  \caption{The ratio $\sigma^{pd}/2\sigma^{pp}$ of Drell-Yan cross
  sections {\em vs.} $x_{2}$.  Shown are next-to-leading order
  calculations, weighted by acceptance, of the Drell-Yan cross section
  ratio using the CTEQ4M (solid) and MRS(R2) (dashed) parton
  distributions.  Also shown is a {\em leading order} calculation of
  the cross section ratio using CTEQ4M (dotted). In the lower CTEQ4M
  curve $\bar{d} - \bar{u}$ has been set to 0 as described in the
  text.  The errors are statistical only.  An additional 1\%
  systematic uncertainty is common to all points.}

  \label{fig:crossratio}
\end{figure}

The resulting ratio of the Drell-Yan cross section per nucleon for
$p+d$ to that for $p+p$ is shown in Fig.~\ref{fig:crossratio} and
Table~\ref{tab:1} as a function of $x_{2}$~\cite{xf,cnb}, the momentum
frac\-tion (Bjorken-$x$) of the target quark in the parton model.
(The Bjorken-$x$ of the beam parton is denoted by $x_1$.)  Muon pairs
with mass, $M_{\mu^+\mu^-}$, below 4.5~GeV/c$^2$ or between 9.0 and
10.7~GeV/c$^2$ were removed to eliminate contributions from the
$J/\psi$ and $\Upsilon$ resonance families.  The data clearly show
that the Drell-Yan cross section per nucleon for $p+d$ exceeds $p+p$
over an appreciable range in $x_{2}$.  Figure~\ref{fig:crossratio}
also shows the predictions for next-to-leading order
calculations~\cite{tung} of the cross section ratio, weighted by the
E866 spectrometer's acceptance, using the CTEQ4M~\cite{lai} and
MRS(R2)~\cite{martin} parton distributions.  The lower curve shows the
predicted ratio for a modified CTEQ4M parton distribution which
maintains the parameterization for $\bar{d}_{p}+\bar{u}_{p}$ but sets
$\bar{d}_{p}-\bar{u}_{p} = 0$.  The data are in reasonable agreement
with the unmodified CTEQ4M and the MRS(R2) predictions for
$x_{2}<0.15$.  It is clear that $\bar{d}_{p} \ne \bar{u}_{p}$ in this
range. Above $x_{2}=0.15$ the data lie well below both
parameterizations.

\begin{table}[tb]

  \caption{Ratio of deuterium to hydrogen cross sections per nucleon
  {\em vs.} $x_2$.  The average of the kinematic variables for each
  bin is also tabulated.  Note that $\langle x_1\rangle = \langle
  x_F\rangle + \langle x_2\rangle$.  The errors are statistical only.
  An additional 1\% systematic uncertainty is common to all
  points.}

  \label{tab:1}
  \begin{tabular}{cccdc}
 & & $\langle p_T\rangle$ & $\langle M_{\mu^+\mu^-}\rangle$ & \\
$\langle x_2\rangle$ & $\langle x_{\text{F}}\rangle$ & (GeV/c)
 &  (GeV/c$^2$)          & $\sigma^{pd}/2\sigma^{pp}$\\ \hline
0.036 & 0.537 & 0.92 & 5.5 & 1.039 $\pm$ 0.017 \\ 
0.057 & 0.441 & 1.03 & 6.5 & 1.079 $\pm$ 0.013 \\ 
0.082 & 0.369 & 1.13 & 7.4 & 1.113 $\pm$ 0.015 \\ 
0.106 & 0.294 & 1.18 & 7.9 & 1.133 $\pm$ 0.020 \\ 
0.132 & 0.244 & 1.21 & 8.5 & 1.196 $\pm$ 0.029 \\ 
0.156 & 0.220 & 1.21 & 9.3 & 1.124 $\pm$ 0.035 \\ 
0.182 & 0.192 & 1.20 & 9.9 & 1.091 $\pm$ 0.043 \\ 
0.207 & 0.166 & 1.19 & 10.6& 1.098 $\pm$ 0.055 \\ 
0.231 & 0.134 & 1.18 & 11.1& 1.055 $\pm$ 0.067 \\ 
0.264 & 0.095 & 1.18 & 11.8& 0.967 $\pm$ 0.067 \\ 
0.312 & 0.044 & 1.12 & 12.8& 0.881 $\pm$ 0.141 \\ 
  \end{tabular}
\end{table}

The acceptance of the spectrometer was largest for $x_{\text{F}} =
x_{1} - x_{2} > 0$.  In this kinematic regime the Drell-Yan cross
section is dominated by the annihilation of a beam quark with a target
antiquark. This fact, coupled with the assumption of charge symmetry
between the neutron and proton and the assumption that the deuteron
parton distributions can be expressed as the sum of the proton and
neutron distributions, yields a simple approximate form of the
Drell-Yan cross section ratio,
\begin{equation} 
\left. \frac{\sigma^{pd}}
           {2\sigma^{pp}}
\right|_{x_1\gg x_2} \approx\frac{1}{2}
\frac{\left( 1 + \frac{1}{4}\frac{d_1}{u_1} \right)}
     {\left( 1 + \frac{1}{4}\frac{d_1}{u_1}
           \frac{\bar{d}_2}{\bar{u}_2} \right)}
      \left( 1 + \frac{\bar{d}_2}{\bar{u}_2} \right).
\label{eq:4}
\end{equation}
The subscripts 1 and 2 denote that the parton distributions in the
proton as functions of $x_1$ and $x_2$, respectively.  In the case
that $\bar{d}=\bar{u}$, the ratio is 1.  This equation illustrates the
sensitivity of the Drell-Yan measurement to $\bar{d}/\bar{u}$ and is
valid only for $x_1\gg x_2$\@.  It does, however, imply an excess of
$\bar{d}$ with respect to $\bar{u}$ for the data.  Estimates of the
nuclear effects in deuterium are significantly less than statistical
errors shown in Fig.~\ref{fig:crossratio}~\cite{e772}.

\begin{figure}
  \begin{center}
    \mbox{\epsffile{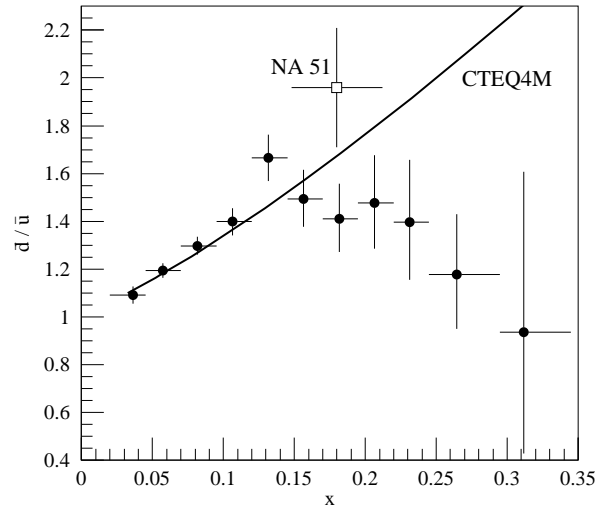}}
  \end{center}

  \caption{The ratio of $\bar{d}/\bar{u}$ in the proton as a function
  of $x$ extracted from the Fermilab E866 cross section ratio. The
  curve is from the CTEQ4M parton distributions.  The error bars
  indicate statistical errors only.  An additional systematic
  uncertainty of $\pm0.032$ is not shown.  The result from NA51 is
  also plotted as an open box.}

  \label{fig:dbub}
\end{figure}

Some of the data, especially at higher $x_2$, do not satisfy the $x_1
\gg x_2$ criterion of Eq.~\ref{eq:4}.  Consequently, $\bar{d}/\bar{u}$
was extracted iteratively by calculating the leading order Drell-Yan
cross section ratio using a set of parton distribution functions as
input and adjusting $\bar{d}/\bar{u}$ until the calculated cross
section ratio agreed with the measured value.  In this procedure, the
values for the $\bar{d}+\bar{u}$, valence and heavy quark
distributions given by the global fits [e.g., CTEQ4M and MRS(R2)] were
assumed to be correct.  In the beam proton, when $x_1 \le 0.345$,
the $\bar{d}/\bar{u}$ distribution was assumed to be the same as in
the target proton.  For $x_1 > 0.345$, a constant value of 1 for
$\bar{d}/\bar{u}$ was used.  Varying the high-$x_1$ value of
$\bar{d}/\bar{u}$ produced almost no change in the low $x_2$ bins, and
less than a 3\% change in the highest $x_2$ bin.  This procedure was
followed using both the CTEQ4M and MRS(R2) parameterizations and
negligible differences were seen.  The extracted $\bar{d}/\bar{u}$
ratio is shown in Fig.~\ref{fig:dbub} along with the prediction made
by CTEQ4M\@.
A qualitative feature of the data, not seen in either
parameterization, is the rapid decrease towards unity of the
$\bar{d}/\bar{u}$ ratio beyond $x=0.2$\@.  At $x = 0.18$, the
extracted $\bar{d}/\bar{u}$ ratio is somewhat smaller than the value
obtained by NA51.  Although the average value of $Q^2$
($M_{\mu^+\mu^-}^2$) is different for the two data sets, the change in
$\bar{d}/\bar{u}$ predicted by the parton distributions due to $Q^2$
evolution is small.

\begin{figure}
  \begin{center}
    \mbox{\epsffile{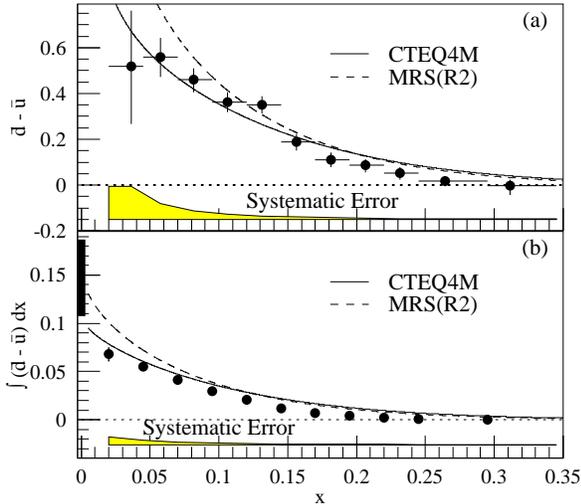}}
  \end{center}

  \caption{Fermilab E866 values for (a) $\bar{d}-\bar{u}$ and (b)
  $\int^{0.345}_{x}\left(\bar{d}-\bar{u}\right) dx'$ in the proton
  versus $x$.  The curves represent the corresponding values
  obtained by the CTEQ4M (solid) and MRS(R2) (dashed)
  parameterizations.  The bar at $0.147\pm0.039$ on the left axis in
  (b) shows the result obtained by NMC for the integral from 0 to 1.}

  \label{fig:d_minus_u}
\end{figure}

To address the GSR violation observed by NMC, the extracted
$\bar{d}/\bar{u}$ ratio is used together with the CTEQ4M value of
$\bar{d} + \bar{u}$ to obtain $\bar{d}-\bar{u}$\@.  (Nearly identical
results are obtained if MRS(R2) is used instead of CTEQ4M.)  Since
each bin in $x$ has a different average $Q^2$, $\bar{d}-\bar{u}$ was
scaled to a common $Q$ value of 7.35~GeV/c, the average for the
entire data set.  Based on this, the integral of $\bar{d}-\bar{u}$
between $x^{\text{min}}$ and 0.345 is calculated.  Both $\bar{d}
-\bar{u}$ and $\int^{0.345}_{x^{\text{min}}}\left(\bar{d}
-\bar{u}\right) dx$ are shown in Fig.~\ref{fig:d_minus_u}.  The
integral reaches a value of $0.068\pm 0.007\text{(stat)}\pm
0.008\text{(syst)}$ at $x^{\text{min}}=0.02$.  This may be
compared with the CTEQ4M and MRS(R2) parameterizations which have
values of 0.076 and 0.100, respectively, for the integral over the
same region.  Over the range $10^{-4}<x<1$, CTEQ4M gives a value of
0.108 for the integral, and MRS(R2) gives 0.160.  Above $x =
0.345$, it is unlikely there are significant contributions to the
$\bar{d}-\bar u$ integral since the sea is relatively small in this
region.  Both CTEQ4M and MRS(R2) find this region contributes less
than 0.002 to the total integral.  It is clear, however, that
significant contributions to the integral arise in the unmeasured
region below $x = 0.02$.

Such a large $\bar{d} /\bar{u}$ asymmetry cannot arise from
perturbative effects~\cite{ross}.  Most parameterizations of the
parton distribution functions (e.g. CTEQ, MRS) simply assume a shape
for $\bar{d}-\bar{u}$ that accommodates the NMC and NA51 results.  It
has been suggested~\cite{thomas,henley} that including the effects of
virtual mesons can account for the observed asymmetry and this
appears~\cite{hwang,kumano2,eichten,szczurek} to be at least
qualitatively correct.

In summary, this Letter reports a measurement of the Drell-Yan cross
section ratio per nucleon of $p+d$ to $p+p$.  From this measurement
the asymmetry of the light quark sea in the proton is extracted as a
function of $x$.  A feature of the present result is the reduction in
$\bar{d}/\bar{u}$ for $x > 0.2$.  The current data are in qualitative
agreement with NA51, but with a smaller $\bar{d}/\bar{u}$ ratio at the
single value of $x$ which they measured. Over the range of $x$
covered in this experiment ($0.020<x<0.345$), the integral of
$\bar{d}-\bar{u}$ reaches a value of
$0.068\pm0.007\text{(stat)}\pm0.008\text{(syst)}$.  However,
contributions to the integral arise from the unmeasured region below
$x = 0.020$.

We would like to thank the Fermilab Particle Physics, Beams and
Computing Divisions for their assistance in performing this
experiment.  This work was supported in part by the U.S. Department of
Energy.


\begin{references}
\bibitem{xf} We define $x_F = \left(x_1 - x_2\right) = 2 p_l/\sqrt{s}$,
             where $p_l$ is the longitudinal momenta of the muon pair
             in the center of mass and $s$ is the total center of mass
             energy squared.
\end{references}
\end{document}